\documentclass{article}

\usepackage{url}

\usepackage{enumitem}
\usepackage{tcolorbox}
\tcbset{
  openproblems/.style={
    colback=gray!6,
    colframe=gray!40,
    boxrule=0.4pt,
    arc=1.5pt,
    left=6pt, right=6pt, top=5pt, bottom=5pt,
    fonttitle=\bfseries\small,
    title=Open Problems
  }
}

\usepackage{longtable}
\usepackage{booktabs}
\usepackage{array}
\usepackage{multirow}
\usepackage{ragged2e}
\usepackage{enumitem}
\usepackage{makecell}
\usepackage[table]{xcolor} 

\usepackage{amssymb}       
\newcolumntype{L}[1]{>{\RaggedRight\arraybackslash}p{#1}}

\usepackage{float}

\usepackage{booktabs}   
\usepackage{graphicx}   
\usepackage{amssymb}    

\usepackage{microtype}
\usepackage{graphicx}
\usepackage{subfigure}
\usepackage{booktabs} 

\usepackage{hyperref}

\usepackage[accepted]{icml2026}

\usepackage{amsmath}
\usepackage{amssymb}
\usepackage{mathtools}
\usepackage{amsthm}

\usepackage{xurl}

\usepackage[capitalize,noabbrev]{cleveref}

\theoremstyle{plain}

\theoremstyle{definition}

\theoremstyle{remark}

\usepackage[textsize=tiny]{todonotes}

\icmltitlerunning{Open Problems in AI Incident Governance
}

\begin{document}

\twocolumn[
\icmltitle{Open Problems in AI Incident Governance
}



\icmlsetsymbol{equal}{*}

\begin{icmlauthorlist}
\icmlauthor{Harleen Kaur Sidhu}{independent}
\icmlauthor{Rebecca Scholefield}{independent}
\icmlauthor{Nour Annan}{sorbonne}
\icmlauthor{Kevin Hernandez}{rice}
\icmlauthor{Isabel Nieh Hou}{columbia}
\icmlauthor{Abdulrahman Alshaikhi}{columbia}
\icmlauthor{Ze Shen Chin}{aisl,oxford}
\icmlauthor{Rokas Gipiškis}{aisl,vu}
\end{icmlauthorlist}

\icmlaffiliation{independent}{Independent}
\icmlaffiliation{aisl}{AI Standards Lab}
\icmlaffiliation{oxford}{Oxford Martin AI Governance Initiative}
\icmlaffiliation{vu}{Vilnius University}
\icmlaffiliation{sorbonne}{Sorbonne University}
\icmlaffiliation{rice}{Rice University}
\icmlaffiliation{columbia}{Columbia University}

\icmlcorrespondingauthor{Rokas Gipiškis}{rokas@aistandardslab.org}

\icmlkeywords{AI incident governance, incident reporting, post-deployment monitoring, AI safety}

\vskip 0.3in
]



\printAffiliationsAndNotice{}  

\begin{abstract}
AI systems may produce failures after deployment that pre-deployment safety assessments do not anticipate. Managing these failures requires what we refer to as adequate \textit{AI incident governance}, where having good definitions, taxonomies, monitoring practices, reporting mechanisms, and incident analysis is essential. 
We examine existing frameworks related to AI incident governance by regulatory bodies and independent efforts, and find that while there are frameworks that describe how individual functions can be performed, there is a lack of consistency within the aspects of definitions, classification, monitoring, and reporting. These inconsistencies apply to the types of incident data that is collected and reported, the ways in which they are categorised, and as a result, the depth, representativeness, and accuracy of analysis that can be performed. 
We identify open problems at each stage of the incident governance pipeline, and find that the absence of standardised monitoring and reporting requirements constitutes a significant gap. To address this, we propose a set of principles supported by concrete monitoring guidelines and a reporting template to facilitate their implementation.

\end{abstract}

\section{Introduction}
\label{submission}

Safety assessments conducted prior to the deployment of an artificial intelligence (AI) system, such as model evaluations and red-teaming, allow testing against anticipated failure modes under controlled settings \cite{shevlane2023model}. However, real-world deployment may produce failures that these assessments cannot anticipate, for example, due to emergent behaviours, adversarial attacks, and unanticipated use cases \cite{o2023deployment, shao2025your}. 

Collecting, reporting, and analysing information about these incidents enables the identification of causal factors, improves accountability, and mitigates future risks of recurrence. Therefore, effective incident governance plays an important role in improving safety and reducing harm caused by AI systems \cite{wei2026designing}. Despite this, we found that there is a lack of consistency across the AI incident governance ecosystem, leading to differences in how incidents are defined, categorised, monitored, reported, and analysed. This limits the comparability of individual incidents and, as a result, reduces the effectiveness of analysis and learning across the field. 



In this paper, we analyse how definitions, taxonomies, monitoring practices, reporting mechanisms, and incident analysis play a role in AI incident governance. Section \ref{sec:definitions} examines existing definitions of AI incidents. Section \ref{sec:taxonomies} reviews the taxonomies through which incidents are classified.  Sections \ref{sec:monitoring} and \ref{sec:reporting} address monitoring and reporting respectively, where we find the central challenges: the lack of robust monitoring procedures and  reporting templates. To address these gaps, we survey corporate monitoring policies (Appendix \ref{sec:policies}), propose monitoring and reporting principles  (Appendices \ref{sec:monitoringprinciples} and \ref{sec:reportingprinciples}), operationalise them as monitoring guidelines (Appendix \ref{sec:guidelines}), and propose a reporting template (Appendix \ref{sec:template}). Finally, Section \ref{sec:learning} considers how incident data can support meaningful analysis.

\section{Definitions}
\label{sec:definitions}

The definition of an AI incident has an effect on what gets monitored, reported, classified, and investigated. For instance, incident repositories based on different definitions may capture different types of events, and therefore different information about them. With the rise of AI regulation worldwide, the definition further determines the obligations and liabilities of actors across the AI value chain. “AI incident” is a relatively new term; as recently as 2024 the Organisation for Economic Co-operation and Development (OECD) described it as an “emerging term” \cite{oecd2024defining}. Defining what counts as an AI incident has to take into account fundamental questions of semantics, values, and law \cite{paeth2024lessons}. We first describe the key definitions of AI incidents in the literature and regulations, highlighting their differences and overlaps, and then identify the open research problems that defining AI incidents still poses.

In the development of the AI Incident Database (AIID), \citet{mcgregor2021preventing} frames an AI incident as a situation in which AI systems caused, or very nearly caused, real-world harm. This broad formulation allows for a wide inclusion of incidents. The OECD offers a more delimited definition of an AI incident: “an event, circumstance or series of events where the development, use or malfunction of one or more AI systems directly or indirectly leads to any of the following harms: (a) injury or harm to the health of a person or groups of people; (b) disruption of the management and operation of critical infrastructure; (c) violations of human rights or a breach of obligations under the applicable law intended to protect fundamental, labour and intellectual property rights; (d) harm to property, communities or the environment” \cite{oecd2024defining}. The same definition is used in OECD’s common framework for AI incident reporting, which is intended to serve as a global benchmark for stakeholders across jurisdictions and sectors \cite{oecd2025framework}. 

Current repositories largely inherit these definitions. The AIID uses a refined version of the definition by \citet{mcgregor2021preventing} by more clearly specifying the scope (affected entities): “an alleged harm or near harm event to people, property, or the environment where an AI system is implicated” \cite{aiid_editors_guide}. The OECD’s AI Incidents Monitor (AIM) adopts the OECD definition. Working toward a mandatory reporting regime, \citet{dixon2025ai} also builds on the OECD definition and proposes a set of standardised components for reporting templates: the type of incident, the nature and severity of harm, technical data, the affected entities and individuals, and the surrounding context. In the report, \citet{dixon2025ai} treats “AI incident" as covering both incidents and near misses. The AI, Algorithmic and Automation Incidents and Controversies (AIAAIC) Repository also includes events that can potentially cause harm. It defines AI incident as “a sudden known or unknown event that becomes public, takes the form of a disruption, loss, emergency, or crisis, and causes or potentially causes harm”  \cite{aiaaic_classifications}.

Recent regulation tends to define narrower, harm-related categories rather than the general term. The EU AI Act defines a “serious incident” as “an incident or malfunctioning of an AI system that directly or indirectly leads to any of the following: (a) the death of a person, or serious harm to a person’s health; (b) a serious and irreversible disruption of the management or operation of critical infrastructure; (c) the infringement of obligations under Union law intended to protect fundamental rights; (d) serious harm to property or the environment"  \cite{euaiact}. The definition is primarily based on realised harms and does not incorporate near misses, similar to \citet{oecd2024defining}. The defined term is the operative trigger for the reporting obligations in Article 73, whose notification deadlines scale with the sub-category of harm (2 days for a serious and irreversible disruption of critical infrastructure, 10 days in the event of death, and 15 days otherwise). California's SB 53 defines a “critical safety incident” as any of the following: “(1) Unauthorised access to, modification of, or exfiltration of, the model weights of a frontier model that results in death or bodily injury. (2) Harm resulting from the materialisation of a catastrophic risk. (3) Loss of control of a frontier model causing death or bodily injury. (4) A frontier model that uses deceptive techniques against the frontier developer to subvert the controls or monitoring of its frontier developer outside of the context of an evaluation designed to elicit this behaviour and in a manner that demonstrates materially increased catastrophic risk" \cite{california_sb53}. This definition includes events that increase risk, such as deceptive evasion, that need not be realised harms. As per SB 53, a frontier developer must report a qualifying critical safety incident to the California Office of Emergency Services within 15 days, or to an appropriate authority within 24 hours where it poses an imminent risk of death or serious physical injury, so a clause such as deceptive evasion determines whether a pre-harm event is reportable at all. New York's Responsible AI Safety and Education Act (RAISE) Act, in its final chapter-amended form \cite{ny_raise_act}, adopts California's “critical safety incident” and its definition, departing from an earlier version that tied safety incidents to critical harm thresholds (death or serious injury of 100 or more people, or one billion dollars in damage). None of these regulations defines the general term “incident”.

The definition of an AI incident should serve as a clarification for several aspects, including potential versus actual harm, the scope and severity of harm, and the nature of the event. However, an important difference remains between two groups of AI incident definitions. Most policy-oriented definitions, including those of the OECD and the EU AI Act, define incidents in relation to realised harm; while other broader definitions also include events that have the potential of causing harm even if harm is not realised. 

\begin{tcolorbox}[openproblems]
\small
\begin{enumerate}[leftmargin=*, itemsep=2pt, parsep=0pt, topsep=0pt, label=\textbf{\arabic*.}]

  \item The scope of AI incidents is inconsistent across definitions — some include potential but unrealised harm, while others exclude it. Should definitions be standardised? Where does a realised incident end and a near miss begin, and should events with potential but unrealised harm count as incidents?
  \item How to validate the quality, usefulness and the intended impact of emerging definitions?
  \item When does a single AI incident begin and end, and when the same failure recurs, should repeat occurrences count as one incident or many? When the harm from one model is spread across many people, or across many separate deployments of that model, how should it be counted as an incident? 
  
\end{enumerate}
\end{tcolorbox}

\section{Taxonomies}
\label{sec:taxonomies}

Taxonomies provide the categories and relationships through which incidents are classified and analysed. Their design determines the extent to which incident records can be meaningfully aggregated across systems, jurisdictions, and time periods. This section surveys the taxonomies that are most closely related to AI incidents, organised by what they classify, namely the causes of failure and the resulting harms, together with their use in incident repositories, before turning to the open problems that taxonomy design still poses.

The Goals, Methods, and Failures (GMF) \cite{pittaras2023failure} taxonomy presents a classification system based on high-level AI system goals (e.g. face recognition), methods and technologies used for system implementation (e.g. transformer neural network), and technical failure causes that result in misbehaviour in the applied system (e.g. distributional bias). Each annotation is paired with a confidence modifier (“known" or “potential") to register how firmly a label can be assigned. There are also other taxonomies that focus on causal factors. Perhaps the most prominent example is the MIT AI Risk Repository's causal taxonomy (entity, intentionality, and timing) \cite{slattery2026ai}, itself descended from the taxonomy of pathways to dangerous AI \cite{yampolskiy2016taxonomy}. While these were conceived to classify AI risks and pathways, the former is also used to categorise incidents captured in the AI Incident Tracker \cite{mit_ai_incident_tracker}.

A second family of taxonomies classifies incidents by the harm they produce. The Center for Security and Emerging Technology (CSET)'s AI Harm Framework \cite{hoffmann2023addingstructureaiharm} separates tangible from intangible harm, requiring tangible harm to involve observable injury, loss, or damage, distinguishes harms that have occurred from those that may yet occur, and sorts them into categories such as harm to physical health or safety, financial loss, and human-rights violations. It is used in the AI Incident Database as the “CSETv1" taxonomy \cite{csetv1}. The MIT AI Risk Repository's domain taxonomy \cite{slattery2026ai} is likewise harm-oriented, sorting cases into seven domains and 24 subdomains and extending the language-model risk taxonomy of \citet{weidinger2022taxonomy}. The former is likewise used to categorise incidents captured in the AI Incident Tracker \cite{mit_ai_incident_tracker}.

Repositories utilise multiple taxonomies. The AIID does not designate a single taxonomy, instead hosting several: the CSET harm taxonomy, GMF, and the MIT taxonomies. This is based on the assumption that reasonable parties may classify the same incident differently \cite{mcgregor2021taxonomy}. In practice none has been applied across the full database.

The key difference in these taxonomies is between classifying why an incident occurred (GMF; the MIT causal taxonomy) and what harm it caused (CSET; the MIT domain taxonomy).  Because the available categories overlap and serve different analytical ends, even reasonable parties may classify the same incident differently \cite{mcgregor2021taxonomy, agarwal2024standardised}. More importantly, the resulting plurality differs in kind from the divergence seen among definitions. Inconsistent definitions across jurisdictions create genuine friction, because a definition fixes obligations and must ultimately be reconciled. Multiple taxonomies, by contrast, may be useful as they address different aspects of incidents. They answer different questions and a repository can effectively use several at once \cite{aiid_taxonomies}.

\begin{tcolorbox}[openproblems]
\small
\begin{enumerate}[leftmargin=*, itemsep=2pt, parsep=0pt, topsep=0pt, label=\textbf{\arabic*.}]

  \item Can incident taxonomies be made mutually exclusive and collectively exhaustive without becoming too coarse to be informative?
  \item How can a category set be chosen so that it stays meaningful as the technology changes (e.g. a method- or technology-based taxonomy designed before the transformer era would now classify most incidents as ``other") without continual revision that breaks comparability over time?
  \item Can comprehensive AI risk taxonomies not designed for incidents be adapted to incident classification, and how should the mismatch between forward-looking risk categories and records of realised incidents be resolved?

\end{enumerate}
\end{tcolorbox}

\section{Monitoring}
\label{sec:monitoring}
Monitoring processes enable the detection of when an incident has occurred and facilitate incident response analysis. This section examines what incident monitoring entails, the actors responsible for monitoring, and the core functions that monitoring performs. 

\subsection{Scope} We refer to incident monitoring as the systematic collection of operational and contextual data that enables organisations to (a) detect that an incident has occurred, and (b) provide technical and organisational context needed to understand its root causes \cite{stein2024role, oecd_ai_incidents_methodology, o2023deployment}.

The term \textit{monitoring} is used across AI literature to describe a wide range of activities across the AI lifecycle, including practices such as tracking performance metrics during training and observing model behaviour during red-teaming and pre-deployment testing \cite{yampolskiy2025monitorability}. However, these pre-deployment modes of monitoring are out of scope of this section, which focuses exclusively on capturing information relating to AI incidents \textit{after} an AI system has been deployed into production environments. 

\subsection{Actors} Across major regulations, international standards, and governance frameworks, primary responsibility for AI monitoring is typically allocated to the organisations that develop, provide, or deploy AI systems \cite{EU_2025, ISO_IEC_42001_2023, nist2023rmf}. These actors conduct system-level monitoring of failures and anomalies based on the operational data that they have access to. However, effective monitoring also requires a distributed ecosystem of external inputs, including users, third-party auditors, researchers, and centralised incident-tracking bodies \cite{stein2024role, jones2023keeping, nist2023rmf}. 

Incident monitoring extends beyond operational surveillance of individual AI systems to tracking incidents across the AI ecosystem. Public AI incident databases and regulatory market-surveillance bodies primarily engage in this form of cross-industry monitoring of media reports and public disclosures. These organisations provide additional avenues for incident report submissions, thereby surfacing incidents that may not be visible through provider-led monitoring, and help ensure accountability, even though they do not replace the need for robust operational monitoring by AI providers and deployers \cite{rodrigues2023artificial}.

\subsection{Functions}

\textbf{Detection.} Detection encompasses the identification and analysis of signals that may indicate that an incident has occurred \cite{pascoe2024nist}. For AI systems, incident detection relies on the continuous monitoring of digital infrastructure to identify signals indicative of harm, such as anomalies and threshold-crossing events. Effective detection encompasses initial triage, where false positives are distinguished from genuine incidents, and the assignment of severity levels to route alerts to appropriate incident response processes \cite{yampolskiy2025monitorability}.

The AI ecosystem has yet to develop standardised classification systems for AI incidents, which may limit the comparability of triage processes across providers, sectors, and jurisdictions. Furthermore, mature high-risk domains systematically incorporate near-miss surveillance as a formal component of detection infrastructure \cite{gnoni2017near}. No equivalent widely adopted near-miss surveillance system currently exists for AI incidents. This may prevent the systemic collection of data that enables the identification of hazardous conditions or behaviours.

Some AI incident signals are distributed across multiple organisational boundaries, including partner companies, cloud providers, downstream developers, and applications that integrate model outputs \cite{o2023deployment}. As no single actor has visibility into all potential indicators of harm, comprehensive incident detection requires a multi-actor monitoring ecosystem. 

Providers of AI systems employ multi-layered automated detection. At inference time, general-purpose filters and real-time automated safety classifiers screen inputs and outputs of AI systems for policy-violating or harmful content. Flagged interactions undergo asynchronous monitoring by classifiers that apply more computationally intensive analysis with greater latency tolerance. Offline safety monitors also review aggregated logs on a periodic basis, enabling the identification of failure modes that real-time classifiers may not be able to detect \cite{phuong2026gdm, Williams_Sun_Sekhar_Carroll_Robinson_Kivlichan, stickland2025async}. 

These automated incident detection mechanisms enable the rapid identification of incident signals that would be infeasible to detect manually at scale. However, the increasing complexity of AI systems and the volume of monitoring signals generated by these detection systems introduce operational challenges. High alert volumes require human operators to regularly triage and validate alerts, creating significant overhead. As AI systems and their corresponding monitoring mechanisms scale, maintaining a balance between sensitivity to incidents and the burden imposed by frequent alerts may prove increasingly difficult. 

Additionally, while automated detection based on system-level operational data is necessary, it is not sufficient for the reliable detection of all incidents that occur. Empirical analysis of production incidents in generative AI cloud services found that 38.3\% of incidents were reported by humans rather than automated monitors, reflecting systemic gaps in automated coverage \cite{yan2025empirical}. Some incidents, especially those with diffused effects, may not be detectable by providers and deployers of AI systems. Detecting these incidents is entirely reliant on incident reports made by affected individuals. These incident reports (which differ from the formal incident reporting described in Section \ref{sec:reporting}) range from bug reports made directly to deployers, to police or news reports, which may then be picked up by media sources. Reports of AI-related harm are monitored by public AI incident databases such as OECD’s AIM detect incidents through monitoring reputable media sources for reports of AI-related harm \cite{oecd_ai_incidents_methodology}. The AIAAIC Repository \cite{aiaaic_user_guide} and the AIID also accept public submission of incident reports \cite{aiid_editors_guide}.  The AIID uses a submission leaderboard to increase coverage of AI incidents by encouraging users to submit more reports \cite{mcgregor2021preventing}.

These continuous and diverse monitoring processes help reduce the time during which incidents remain undetected, limiting the opportunity for harms to propagate. Furthermore, improved detection by public AI incident databases enhances the empirical foundations for forecasting and learning, while simultaneously creating external pressure on providers to maintain high monitoring and accountability standards \cite{mcgregor2021preventing, rodrigues2023artificial}. 

\textbf{Preservation of contextual information.} Understanding the causes of an incident requires monitoring that preserves both the technical and organisational context in which the system operated \cite{dixon2025ai}. Technical context includes information about how users interacted with the system as well as actions taken by agents, and changes in task specifications or permissions \cite{anderljung2023frontier, ezell2025incident, eu_gpai_code}. These logs may provide insights into how the system’s behaviour diverged from its intended operation. The organisational context includes post-deployment modifications in design decisions, documented safeguards, risk-assumption registers, oversight structures, and approval or escalation pathways \cite{nist2023rmf}. These records may allow investigators to determine if governance choices or procedural gaps may have contributed to the incident. Together, this data helps to trace the sequence of decisions, interactions, and events that led to the occurrence of an incident \cite{euaiact}. 

To ensure that all relevant contextual evidence is captured and retained, it may be necessary to employ continuous, real-time monitoring mechanisms and processes to track and preserve applicable logs and documents \cite{eu_gpai_code, ferdaus2026towards}. However, some information, such as context windows provided by users, may  not be sufficiently monitorable by deployers \cite{paeth2024lessons}. Preserving sufficient contextual information also poses unique challenges for autonomous and distributed systems, such as multi-agent architectures. Existing memory systems for agentic AI may not sufficiently support traceability and incident investigation \cite{wang2026agenttracestrustsurvey}. Finally, monitoring that involves logging user data raises concerns about privacy \cite{stein2024role}. Even when organisations employ pseudonymization or aggregation methods to protect sensitive data, research on de-anonymisation attacks demonstrates that sensitive information can still be recovered by malicious actors \cite{xin2025false, feretzakis2024trustworthy, lange2025slice}.

\begin{tcolorbox}[openproblems]
\small
\begin{enumerate}[leftmargin=*, itemsep=2pt, parsep=0pt, topsep=0pt, label=\textbf{\arabic*.}]
  \item Should the triaging processes be standardised across the AI ecosystem, and if so, how should it be standardised?
  \item How should different actors share responsibility for AI incident monitoring when incident signals are so fragmented across the AI ecosystem?
  \item How can automated detection systems be kept up to date as models and threat vectors change rapidly over time, and how should they be calibrated to manage trade-offs between detection sensitivity and alert fatigue?
  \item How can monitoring systems balance the need for comprehensive user data with privacy protection?
\end{enumerate}
\end{tcolorbox}

\section{Reporting}
\label{sec:reporting}

The EU AI Act establishes incident reporting requirements \cite{euaiact}, and the European Commission has published draft report templates to support implementation \cite{ec2025template}. This section compares its approach with frameworks from the OECD \cite{oecd2025framework} and CSET \cite{dixon2025ai, dixon2024argument}.

\subsection{Scope}
There is debate about the appropriate scope of mandatory reporting. The EU AI Act requires reporting only serious incidents, where AI systems cause harm, including “the death of a person or serious damage to a person’s health, to property or the environment” \cite{euaiact}. Others argue for mandatory reporting of all incidents, regardless of severity, and near misses \cite{dixon2025ai}. 

Increasing the scope of mandatory reporting may increase the burden on reporters and receiving authorities, particularly if reports require manual review. However, some argue that reporting all incidents strengthens oversight of emerging risks, including “systemic” harms that appear insignificant in isolation but become severe in aggregate \cite{paeth2024lessons}.

\subsection{Actors}
As discussed in Section \ref{sec:definitions}, certain regulations place obligations on incident reporting to public authorities onto those who develop, provide, or deploy AI systems. Separately, some have also argued for supporting members of the public to report incidents voluntarily, including private individuals who experience or observe them, as well as stakeholders such as researchers, journalists, and watchdogs \cite{dixon2024argument}. However, there is limited work on how to support voluntary reporting—for example, how barriers posed by limited technical knowledge can be overcome.

\subsection{Common elements of incident reports}

\textbf{Timelines.} Each framework asks about the timeline of an incident, though details vary. The European Commission distinguishes between start date, end date, and date of detection \cite{ec2025template}. However, these events are not formally defined: “end date” could be interpreted as the date on which a system was fixed, or the date on which harms ceased or were remediated. 
It is also unclear how precise dates should be—for instance, whether a month and year suffice if exact dates are unknown. A related question is how reports should capture uncertainty about dates. One approach is to require reporters to indicate whether dates are known, estimated, or unknown \cite{pittaras2023failure}.

\textbf{Implicated systems.} Frameworks vary in the technical and operational information requested for AI systems that contributed to an incident. The European Commission focuses on identifying the specific product via its EU database ID, serial or batch number, and software and firmware versions \cite{ec2025template}. CSET proposes more detailed technical documentation, including system cards, model cards, and datasheets \cite{dixon2025ai}. The OECD places greater emphasis on deployment context, including usage context and rights, a system’s autonomy, and whether incidents involve multiple interacting systems \cite{oecd2025framework}.

This variation reflects a lack of consensus about the role of incident reports \cite{stein2024role}. Reports could aim to provide enough information for causal analysis, or they could simply identify incidents warranting further investigation. Reports may also enable pattern identification—for example, by flagging whether models lacking certain safeguards, designed for particular use cases, or deployed in certain domains are disproportionately implicated in harm. Each approach calls for different kinds of information.
A further question is how information requirements should differ across multiple interacting systems, and how reports can capture their contributions.

\textbf{Causality.} Proposed frameworks differ substantially in how they approach causality. Because the European Commission frames reporting as an iterative process, it requires relatively little causal information upfront, asking “What went wrong with the system?” and “What is the likely cause?” and deferring definitive root causes to final reports \cite{ec2025template}. The OECD and CSET frameworks are more analytical, attempting to capture causal information through comprehensive categories from the outset. For example, the OECD combines questions about the nature of causality—direct cause or contributing factor—with more specific categories such as overreliance, intentional misuse, and human error, and allows reporters to expand on failures at the data, model, and system levels \cite{oecd2025framework}.

Where reporting is iterative, an open question is how much causal information initial reports should seek. Open-ended questions with insufficient guidance may produce uninformative reports, unless followed by formal investigation. A related question is how to balance the flexibility of open-text fields with the comprehensiveness of structured categories.

\textbf{Impact.} Across frameworks, there is a consistent focus on core harm types—including physical, environmental, and financial harms—and each framework attempts to quantify their severity. Frameworks differ mainly in how many dimensions they propose: CSET, for instance, adds remediability, optionality, and frequency \cite{dixon2025ai}. In terms of who is impacted, the European Commission focuses narrowly on users, while the OECD covers a broader range of stakeholders, asking reporters to specify whether groups such as ``children", ``trade unions", or ``businesses" were impacted \cite{oecd2025framework}. 

An open question is what should be mandatory: some fields—such as the number of users harmed or the remediability of an incident—may be difficult to assess at the point of reporting and may be better treated as optional or follow-up submissions.

\begin{tcolorbox}[openproblems]
\small
\begin{enumerate}[leftmargin=*, itemsep=2pt, parsep=0pt, topsep=0pt, label=\textbf{\arabic*.}]
  \item What purposes should incident reports serve, and what information is required as a result?
  \item What should be the scope of mandatory incident reporting, and how should voluntary reporting be supported outside this scope?
  \item How should information requirements vary across stages of an iterative reporting process?
  \item How should reports balance structured categories with open-ended reporting?
\end{enumerate}
\end{tcolorbox}

\section{Incident analysis}
\label{sec:learning}

Establishing how AI incidents are defined and taxonomised standardises the types of events that are recognised as incidents and the ways in which they are categorised. The need for incidents to be meaningfully analysed and understood, in turn, determines what needs to be monitored and reported. However, the mere collection of AI incident information does not, in itself, enable learning and reduce risk \cite{turri2023we}. The main goal of incident analysis is to prevent future incidents from occurring, including through improved organisational practices. In this section, we describe single-incident causal analysis and aggregate cross-incident analysis, and then identify the open research problems in AI incident analysis.

Given one event, single-incident causal analysis reconstructs what happened and why. It involves identifying the technical, organisational, and contextual factors that lead to AI incidents. \citet{mylius2024rootcause} demonstrates fault-tree analysis of AI safety incidents, using language models to infer candidate causes from incident reports. \citet{ezell2025incident} introduce a causal framework for AI-agent incidents that, drawing on human-factors methods, traces failures through system factors (e.g.\ training data and learning methods), contextual factors (e.g.\ prompt injections and tool access), and cognitive errors (e.g.\ misinterpreting a request), while specifying the activity logs and system information an investigator needs to determine which factors apply. 

Given a corpus of incidents, aggregate cross-incident analysis may help in finding patterns, identifying recurring contributors, and understanding emerging risks. Beyond understanding incidents at an individual level, analysis on a population level requires a common taxonomy and reporting system, but can provide insights on broader trends. Analyses of incidents can enable a more comprehensive understanding of historical trends and the current state of the world. Large-scale analyses of public AI incident databases have been used to examine recurring harm categories \cite{may2024sok, velazquez2024decoding}, identify gaps between developer risk assumptions and real-world impacts \cite{rao2025ai}, and analyse patterns in accountability and response following AI incidents \cite{richards2025accountability}. 

Additionally, extrapolation of historical trends can also lead to insights into future trends. Forecasting extends pattern identification toward estimating the frequency of future incidents. Incident analysis remains nascent in AI safety, with existing work primarily centred around generalising and causal analysis. However, incident forecasting is an established practice in other safety-critical domains \cite{zheng2009overview}. It has applications in epidemiology \cite{desai2019real, rilkoff2024innovations, cdc_nowcasting}, aviation \cite{zhang2019ensemble}, cybersecurity \cite{almahmoud2023holistic}, and the maritime domain \cite{kandel2024data}, where it can be used to identify frequent incident types and predict future incident occurrences for them.

\begin{tcolorbox}[openproblems]
\small
\begin{enumerate}[leftmargin=*, itemsep=2pt, parsep=0pt, topsep=0pt, label=\textbf{\arabic*.}]
  \item How can incident datasets balance high-dimensional categorical fields with free-text descriptions to enable both rich contextual detail and systematic analysis at scale?
  \item How can forecasting approaches identify indicators of emerging risks, particularly for novel failure modes?
  \item What forecasting methodologies from other safety-critical domains can be adapted to AI incident governance given the specific characteristics of AI systems?
\end{enumerate}
\end{tcolorbox}

\section{Conclusion}

AI incident governance is central to ensuring a safe and secure AI ecosystem, yet existing frameworks lack consistency across key aspects of definitions, taxonomies, monitoring and reporting practices, and incident analyses. As AI is increasingly deployed in high-risk contexts and the risk landscape evolves, robust incident governance processes are key. 
Effective AI incident governance depends on definitions that clearly establish scope and responsibilities, comprehensive taxonomies enabling causal analyses and categorization of harm, and monitoring and reporting practices that capture and document sufficient data for incident analysis.

This paper has identified key open problems at each stage of the incident governance pipeline, most notably the lack of robust monitoring and reporting practices. The monitoring and reporting principles, monitoring guidelines, and reporting template proposed in the appendix represent an initial response to the most pressing operational gaps. Further research is also needed on topics such as forecasting methodologies, and the harmonisation of reporting obligations across jurisdictions.

\section*{Impact Statement}

This paper contributes to ongoing efforts to make AI deployment more accountable by surfacing the gaps between how incidents are defined, monitored, and reported. The proposed monitoring guidelines and reporting template are intended as starting points for harmonisation across frameworks, not as final specifications. 

\section*{Use of Large Language Models}
During the preparation of this work, large language models (Claude, Gemini) were used to conduct preliminary literature exploration and to edit and review drafts. All AI-generated text and suggested sources were manually reviewed, verified, and edited by the authors, who take full responsibility for the final content of the paper.

\section*{Acknowledgements}

The authors thank the SPAR (Supervised Program for Alignment Research) AI safety program for supporting this work. We also thank Marcel Mir Teijeiro, Koen Holtman, Sean McGregor, George Gor, and Adrian Regenfuß for their comments on earlier versions of this draft. Any remaining errors are our own.



\bibliography{example_paper}

@report{oecd2025framework,
  title        = {{Towards a Common Reporting Framework for AI Incidents: The 29 Criteria}},
  author       = {{OECD}},
  year         = {2025},
  institution  = {OECD Publishing},
  url          = {https://www.oecd.org/content/dam/oecd/en/publications/reports/2025/02/towards-a-common-reporting-framework-for-ai-incidents_8c488fdb/f326d4ac-en.pdf}
}

@report{oecd2024defining,
  title        = {{Defining AI Incidents and Related Terms}},
  author       = {{OECD}},
  year         = {2024},
  institution  = {OECD Publishing}
}

@inproceedings{weidinger2022taxonomy,
  title={{Taxonomy of risks posed by language models}},
  author={Weidinger, Laura and Uesato, Jonathan and Rauh, Maribeth and Griffin, Conor and Huang, Po-Sen and Mellor, John and Glaese, Amelia and Cheng, Myra and Balle, Borja and Kasirzadeh, Atoosa and others},
  booktitle={Proceedings of the 2022 ACM conference on fairness, accountability, and transparency},
  pages={214--229},
  year={2022}
}

@inproceedings{agarwal2024standardised,
  title={{Standardised schema and taxonomy for AI incident databases in critical digital infrastructure}},
  author={Agarwal, Avinash and Nene, Manisha J},
  booktitle={2024 IEEE Pune Section International Conference (PuneCon)},
  pages={1--6},
  year={2024},
  organization={IEEE}
}

@misc{richards2025accountability,
  author = {Richards, I. and Benn, C. and Zilka, M.},
  title = {{From Incidents to Insights: Patterns of Responsibility following {AI} Harms}},
  year = {2025},
  eprint = {2505.04291},
  archivePrefix = {arXiv},
  primaryClass = {cs.CY}
}

@misc{paeth2024lessons,
  author = {Paeth, K. and Atherton, D. and Pittaras, N. and Frase, H. and McGregor, S.},
  title = {{Lessons for Editors of {AI} Incidents from the {AI} Incident Database}},
  year = {2024},
  eprint = {2409.16425},
  archivePrefix = {arXiv},
  primaryClass = {cs.CY}
}

@misc{aiaaic_user_guide, 
author={AIAAIC},
title={{AIAAIC Repository user guide}}, 
howpublished ={\url{https://www.aiaaic.org/aiaaic-repository/user-guide}}, 
year={2024}, 
month={May},
note = {[Accessed 06-11-2025]}}

@inproceedings{ezell2025incident,
  title={{Incident Analysis for AI Agents}},
  author={Ezell, Carson and Roberts-Gaal, Xavier and Chan, Alan},
  booktitle={Proceedings of the AAAI/ACM Conference on AI, Ethics, and Society},
  volume={8},
  number={1},
  pages={865--878},
  year={2025}
}

@article{yampolskiy2025monitorability,
  title={{On monitorability of AI}},
  author={Yampolskiy, Roman V},
  journal={AI and Ethics},
  volume={5},
  number={1},
  pages={689--707},
  year={2025},
  publisher={Springer}
}

@article{buhl2024safety,
  title={Safety cases for frontier {AI}},
  author={Buhl, Marie Davidsen and Sett, Gaurav and Koessler, Leonie and Schuett, Jonas and Anderljung, Markus},
  journal={arXiv preprint arXiv:2410.21572},
  year={2024}
}

@article{bluemke2023exploring,
  title={{Exploring the relevance of data Privacy-Enhancing technologies for AI governance use cases}},
  author={Bluemke, Emma and Collins, Tantum and Garfinkel, Ben and Trask, Andrew},
  journal={arXiv preprint arXiv:2303.08956},
  year={2023}
}

@article{o2023deployment,
  title={{Deployment corrections: An incident response framework for frontier {AI} models}},
  author={O'Brien, Joe and Ee, Shaun and Williams, Zoe},
  journal={arXiv preprint arXiv:2310.00328},
  year={2023}
}

@article{shahriar2023survey,
  title={{A survey of privacy risks and mitigation strategies in the artificial intelligence life cycle}},
  author={Shahriar, Sakib and Allana, Sonal and Hazratifard, Seyed Mehdi and Dara, Rozita},
  journal={IEEE Access},
  volume={11},
  pages={61829--61854},
  year={2023},
  publisher={IEEE}
}

@inproceedings{turri2023we,
  title={{Why we need to know more: Exploring the state of AI incident documentation practices}},
  author={Turri, Violet and Dzombak, Rachel},
  booktitle={Proceedings of the 2023 AAAI/ACM Conference on AI, Ethics, and Society},
  pages={576--583},
  year={2023}
}

@article{dixon2024argument,
  title={An argument for hybrid {AI} incident reporting},
  author={Dixon, Ren Bin Lee and Frase, Heather},
  journal={Center for Security and Emerging Technology. Retrieved from https://cset. georgetown. edu/publication/an-argument-for-hybrid-ai-incident-reporting/(accessed 2024-06-04)},
  year={2024}
}

@article{dixon2025ai,
  title={{AI} incidents: Key components for a mandatory reporting regime},
  author={Dixon, Ren Bin Lee and Frase, Heather},
  journal={Georgetown Center for Security and Emerging Technology},
  year={2025}
}

@misc{EU_2025, 
author={{European Commission}},
title={{Incident report for serious incidents under the {AI} Act (High-risk AI systems)}}, 
howpublished={\url{https://digital-strategy.ec.europa.eu/en/consultations/ai-act-commission-issues-draft-guidance-and-reporting-template-serious-ai-incidents-and-seeks}},
year={2025},
month={September},
note = {[Accessed 06-11-2025]}}

@misc{ec2025template,
  title        = {{AI Act: Commission publishes a reporting template for serious incidents involving general-purpose AI models with systemic risk}},
  author       = {{European Commission}},
  year         = {2025},
  howpublished = {\url{https://digital-strategy.ec.europa.eu/en/library/ai-act-commission-publishes-reporting-template-serious-incidents-involving-general-purpose-ai}}
}

@techreport{nist2023rmf,
  title        = {{Artificial Intelligence Risk Management Framework (AI RMF 1.0)}},
  author       = {{NIST}},
  year         = {2023},
  institution  = {U.S. Department of Commerce},
  url          = {https://nvlpubs.nist.gov/nistpubs/ai/nist.ai.100-1.pdf}
}

@inproceedings{pittaras2023failure,
  title        = {{A taxonomic system for failure cause analysis of open source AI incidents}},
  author       = {Pittaras, Nikolaos and McGregor, Sean},
  booktitle    = {Proceedings of the SafeAI 2023 Workshop},
  year         = {2023},
  url          = {https://ceur-ws.org/Vol-3381/17.pdf}
}

@article{stein2024role,
  title={The role of governments in increasing interconnected post-deployment monitoring of {AI}},
  author={Stein, Merlin and Bernardi, Jamie and Dunlop, Connor},
  journal={arXiv preprint arXiv:2410.04931},
  year={2024}
}

@misc{euaiact,
  
  title = {Regulation ({EU}) 2024/1689 of the {European Parliament} and of the {Council} of 13 {June} 2024 laying down harmonised rules on artificial intelligence...},
  author = {{European Parliament} and {Council of the European Union}},
  shortauthor = {{Regulation (EU) 2024/1689}},
  year = {2024},
  journal = {Official Journal of the European Union},
  volume = {L 2024/1689},
  url = {https://eur-lex.europa.eu/eli/reg/2024/1689/oj/eng},
  howpublished = {\url{https://eur-lex.europa.eu/eli/reg/2024/1689/oj/eng}},
   note         = {Official Journal of the European Union, OJ L 2024/1689, 12 July 2024. Text with EEA relevance},
}

@misc{ISO_IEC_42001_2023,
  shorthand = {{ISO/IEC 42001:2023}},
  author = {{ISO/IEC}},
  title        = {{ISO/IEC 42001:2023 Information technology --- Artificial intelligence --- Management system}},
  year         = {2023},
  howpublished = {\url{https://www.iso.org/standard/81230.html}},
  note         = {ISO/IEC 42001:2023 Artificial Intelligence Management System ({AIMS}) standard},
}

@report{jones2023keeping,
  author       = {Jones, Elliot and Birtwistle, Michael and Reid, Octavia Field},
  title        = {Keeping an Eye on {AI}: Approaches to Government Monitoring of the {AI} Landscape},
  institution  = {Ada Lovelace Institute},
  year         = {2023},
  month        = {July},
  url          = {https://www.adalovelaceinstitute.org/report/keeping-an-eye-on-ai/},
  note         = {Accessed: 2026-02-14}
}

@article{rodrigues2023artificial,
  title={When artificial intelligence fails: The emerging role of incident databases},
  author={Rodrigues, Rowena and Resseguier, Anais and Santiago, Nicole},
  journal={Pub. Governance, Admin. \& Fin. L. Rev.},
  volume={8},
  pages={17},
  year={2023},
  publisher={HeinOnline}
}

@misc{eu_gpai_code,
  author       = {{European Commission}},
  title        = {{The General-Purpose {AI} Code of Practice}},
  year         = {2025},
  howpublished = {\url{https://digital-strategy.ec.europa.eu/en/policies/contents-code-gpai}},
  note         = {Accessed: 2026-02-14}
}

@inproceedings{mcgregor2021preventing,
  title={{Preventing repeated real world {AI} failures by cataloging incidents: The {AI} incident database}},
  author={McGregor, Sean},
  booktitle={Proceedings of the {AAAI} Conference on Artificial Intelligence},
  volume={35},
  number={17},
  pages={15458--15463},
  year={2021}
}

@misc{oecd_ai_incidents_methodology,
  author       = {{OECD}},
  title        = {Overview and methodology of the {AI} Incidents and Hazards Monitor
Methodology and disclosures},
  howpublished = {\url{https://oecd.ai/en/incidents-methodology}},
  note         = {Accessed: 2026-02-14}
}

@misc{aiid_editors_guide,
  author       = {{{AI} Incident Database}},
  title        = {Editors' Guide},
  howpublished = {\url{https://incidentdatabase.ai/editors-guide/}},
  note         = {Accessed: 2026-02-14}
}

@article{anderljung2023frontier,
  title={{Frontier {AI} regulation: Managing emerging risks to public safety}},
  author={Anderljung, Markus and Barnhart, Joslyn and Korinek, Anton and Leung, Jade and O'Keefe, Cullen and Whittlestone, Jess and Avin, Shahar and Brundage, Miles and Bullock, Justin and Cass-Beggs, Duncan and others},
  journal={arXiv preprint arXiv:2307.03718},
  year={2023}
}

@article{ferdaus2026towards,
  title={Towards trustworthy {AI: A} review of ethical and robust large language models},
  author={Ferdaus, Md Meftahul and Abdelguerfi, Mahdi and Loup, Elias and N. Niles, Kendall and Pathak, Ken and Sloan, Steven},
  journal={ACM Computing Surveys},
  volume={58},
  number={7},
  pages={1--43},
  year={2026},
  publisher={ACM New York, NY}
}

@article{scheuerman2021framework,
  title={A framework of severity for harmful content online},
  author={Scheuerman, Morgan Klaus and Jiang, Jialun Aaron and Fiesler, Casey and Brubaker, Jed R},
  journal={Proceedings of the ACM on Human-Computer Interaction},
  volume={5},
  number={CSCW2},
  pages={1--33},
  year={2021},
  publisher={ACM New York, NY, USA}
}

@techreport{hoffmann2023addingstructureaiharm,
  title        = {{Adding Structure to {AI H}arm: {A}n {I}ntroduction to {CSET}'s {AI H}arm {F}ramework}},
  author       = {Hoffmann, Mia and Frase, Heather},
  institution  = {CSET (CSET), Georgetown University},
  year         = {2023},
  month        = {July},
  doi          = {10.51593/20230022},
  url          = {https://cset.georgetown.edu/publication/adding-structure-to-ai-harm/}
}

@misc{mylius2023aiharmseverity,
  title        = {{AI Harm Severity Scales by Category}},
  author       = {Mylius, Simon},
  howpublished = {\url{https://simonmylius.com/ai-harm-severity-scales}},
  note         = {Accessed: 2026-02-14}
}

@article{xin2025false,
  title={A false sense of privacy: Evaluating textual data sanitization beyond surface-level privacy leakage},
  author={Xin, Rui and Mireshghallah, Niloofar and Li, Shuyue Stella and Duan, Michael and Kim, Hyunwoo and Choi, Yejin and Tsvetkov, Yulia and Oh, Sewoong and Koh, Pang Wei},
  journal={arXiv preprint arXiv:2504.21035},
  year={2025}
}

@article{feretzakis2024trustworthy,
  title={{Trustworthy AI: Securing sensitive data in large language models}},
  author={Feretzakis, Georgios and Verykios, Vassilios S},
  journal={AI},
  volume={5},
  number={4},
  pages={2773--2800},
  year={2024},
  publisher={MDPI}
}

@inproceedings{lange2025slice,
  title={Slice it up: Unmasking user identities in smartwatch health data},
  author={Lange, Lucas and Schreieder, Tobias and Christen, Victor and Rahm, Erhard},
  booktitle={Proceedings of the 20th ACM Asia Conference on Computer and Communications Security},
  pages={710--726},
  year={2025}
}

@article{szadeczky2025risk,
  title={Risk, regulation, and governance: evaluating artificial intelligence across diverse application scenarios: Risk, regulation, and governance},
  author={Szadeczky, Tamas and Bederna, Zsolt},
  journal={Security Journal},
  volume={38},
  number={1},
  pages={35},
  year={2025},
  publisher={Springer}
}

@article{johnson2024developing,
  title={Developing real-time monitoring models to enhance operational support and improve incident response times},
  author={Johnson, Omoniyi Babatunde and Olamijuwon, Jeremiah and Cadet, Emmanuel and Osundare, Olajide Soji and Weldegeorgise, Yodit Wondaferew},
  journal={Int J Eng Res Dev},
  volume={20},
  number={11},
  pages={1296--1304},
  year={2024}
}

@article{tariq2025alert,
  title={Alert fatigue in security operations centres: Research challenges and opportunities},
  author={Tariq, Shahroz and Baruwal Chhetri, Mohan and Nepal, Surya and Paris, Cecile},
  journal={ACM Computing Surveys},
  volume={57},
  number={9},
  pages={1--38},
  year={2025},
  publisher={ACM New York, NY}
}

@article{vei2025ai,
  title={{AI Harmonics: a human-centric and harms severity-adaptive AI risk assessment framework}},
  author={Vei, Sofia and Giudici, Paolo and Sermpezis, Pavlos and Vakali, Athena and Bernardelli, Adelaide Emma},
  journal={arXiv preprint arXiv:2509.10104},
  year={2025}
}

@article{vinnakotacreating,
  title={Creating Effective Alerts for Monitoring Distributed Systems},
  author={Vinnakota, Krishna and Kolla, Madhuri},
  journal={International Journal of Computer Trends and Technology},
  volume={73},
  pages={172--178},
  year={2025}
}

@standard{iso_iec_22989_2022,
  author = {{ISO/IEC}},
  shorthand = {{ISO/IEC 22989:2022}},
  title        = {{ISO/IEC 22989:2022 Information technology — Artificial intelligence — Artificial intelligence concepts and terminology}},
  organization = {{International Organization for Standardization}},
  year         = {2022},
}

@standard{iso_iec_23053_2022,
  key = {{ISO/IEC 23054:2022}},
  author = {{ISO/IEC}},
  title        = {{ISO/IEC 23054:2022} Framework for Artificial Intelligence (AI) systems using Machine Learning (ML)},
  organization = {{International Organization for Standardization}},
  year         = {2022},
}

@standard{iso_iec_23894_2023,
   key = {{ISO/IEC 23894:2023}},
  author = {{ISO/IEC}},
  title        = {{ISO/IEC 23894:2023} Information technology — Artificial intelligence — Guidance on risk management},
  organization = {{International Organization for Standardization}},
  year         = {2023},
}

@standard{iso_iec_27035-1_2023,
  key = {{ISO/IEC 27035:2023}},
  author = {{ISO/IEC}},
  title        = {{ISO/IEC 27035-1:2023}Information technology — Information security incident management},
  organization = {{International Organization for Standardization}},
  year         = {2023}
}

@article{shevlane2023model,
  title={Model evaluation for extreme risks},
  author={Shevlane, Toby and Farquhar, Sebastian and Garfinkel, Ben and Phuong, Mary and Whittlestone, Jess and Leung, Jade and Kokotajlo, Daniel and Marchal, Nahema and Anderljung, Markus and Kolt, Noam and others},
  journal={arXiv preprint arXiv:2305.15324},
  year={2023}
}

@misc{mylius2024rootcause,
  author = {Mylius, Simon},
  title = {Root Cause Analysis of {AI} Safety Incidents},
  year = {2024},
  month = jun,
  url = {{https://simonmylius.com/blog/6wj3yx02hivp2vbl1bz9dqmxmelc35}},
  note = {Blog post},
  urldate = {2026-02-16}
}

@article{zhang2019ensemble,
  title={Ensemble machine learning models for aviation incident risk prediction},
  author={Zhang, Xiaoge and Mahadevan, Sankaran},
  journal={Decision Support Systems},
  volume={116},
  pages={48--63},
  year={2019},
  publisher={Elsevier}
}

@inproceedings{may2024sok,
  title={Sok: How artificial-intelligence incidents can jeopardize safety and security},
  author={May, Richard and Kr{\"u}ger, Jacob and Leich, Thomas},
  booktitle={Proceedings of the 19th international conference on availability, reliability and security},
  pages={1--12},
  year={2024}
}

@article{velazquez2024decoding,
  title={Decoding real-world artificial intelligence incidents},
  author={Vel{\'a}zquez, Julia De Miguel and {\v{S}}{\'c}epanovi{\'c}, Sanja and Gvirtz, Andr{\'e}s and Quercia, Daniele},
  journal={Computer},
  volume={57},
  number={11},
  pages={71--81},
  year={2024},
  publisher={IEEE}
}

@inproceedings{rao2025ai,
  title={{The AI Model Risk Catalog: What Developers and Researchers Miss About Real-World AI Harms}},
  author={Rao, Pooja SB and {\v{S}}{\'c}epanovi{\'c}, Sanja and Jayagopi, Dinesh Babu and Cherubini, Mauro and Quercia, Daniele},
  booktitle={{Proceedings of the AAAI/ACM Conference on AI, Ethics, and Society}},
  volume={8},
  number={3},
  pages={2150--2163},
  year={2025}
}

@article{zheng2009overview,
  title={An overview of accident forecasting methodologies},
  author={Zheng, Xiaoping and Liu, Mengting},
  journal={Journal of Loss Prevention in the process Industries},
  volume={22},
  number={4},
  pages={484--491},
  year={2009},
  publisher={Elsevier}
}

@article{desai2019real,
  title={Real-time epidemic forecasting: challenges and opportunities},
  author={Desai, Angel N and Kraemer, Moritz UG and Bhatia, Sangeeta and Cori, Anne and Nouvellet, Pierre and Herringer, Mark and Cohn, Emily L and Carrion, Malwina and Brownstein, John S and Madoff, Lawrence C and others},
  journal={Health security},
  volume={17},
  number={4},
  pages={268--275},
  year={2019},
  publisher={SAGE Publications Sage CA: Los Angeles, CA}
}

@article{rilkoff2024innovations,
  title={Innovations in public health surveillance: An overview of novel use of data and analytic methods},
  author={Rilkoff, Heather and Struck, Shannon and Ziegler, Chelsea and Faye, Laura and Paquette, Dana and Buckeridge, David},
  journal={Canada Communicable Disease Report},
  volume={50},
  number={3-4},
  pages={93},
  year={2024}
}

@misc{cdc_nowcasting,
  author = {{Centers for Disease Control and Prevention}},
  title = {Behind the Model: Nowcasting},
  howpublished = {\url{https://www.cdc.gov/cfa-behind-the-model/php/data-research/nowcasting.html}},
  note = {Accessed: 2026-04-25},
  year = {2026}
}

@article{kandel2024data,
  title={A data-driven risk assessment of Arctic maritime incidents: Using machine learning to predict incident types and identify risk factors},
  author={Kandel, Rajesh and Baroud, Hiba},
  journal={Reliability Engineering \& System Safety},
  volume={243},
  pages={109779},
  year={2024},
  publisher={Elsevier}
}

@article{almahmoud2023holistic,
  title={A holistic and proactive approach to forecasting cyber threats},
  author={Almahmoud, Zaid and Yoo, Paul D and Alhussein, Omar and Farhat, Ilyas and Damiani, Ernesto},
  journal={Scientific Reports},
  volume={13},
  number={1},
  pages={8049},
  year={2023},
  publisher={Nature Publishing Group UK London}
}

@article{pascoe2024nist,
  title={{The NIST cybersecurity framework (CSF) 2.0}},
  author={Pascoe, Cherilyn and Quinn, Stephen and Scarfone, Karen},
  year={2024},
  publisher={Cherilyn Pascoe, Stephen Quinn, Karen Scarfone}
}

@misc{wang2026agenttracestrustsurvey,
      title={{From Agent Traces to Trust: A Survey of Evidence Tracing and Execution Provenance in LLM Agents}}, 
      author={Yiqi Wang and Jiaqi Zhang and Taotao Cai and Zirui Liu and Qingqiang Sun and Zequn Sun and Zhangkai Wu and Manqing Dong and Mingkai Zhang and Xuefei Yin and Yanming Zhu},
      year={2026},
      eprint={2606.04990},
      archivePrefix={arXiv},
      primaryClass={cs.CR},
      url={https://arxiv.org/abs/2606.04990}, 
}

@inproceedings{yan2025empirical,
  title={{An empirical study of production incidents in generative AI cloud services}},
  author={Yan, Haoran and Chen, Yinfang and Ma, Minghua and Wen, Ming and Lu, Shan and Zhang, Shenglin and Xu, Tianyin and Wang, Rujia and Bansal, Chetan and Rajmohan, Saravan and others},
  booktitle={2025 IEEE 36th International Symposium on Software Reliability Engineering (ISSRE)},
  pages={359--370},
  year={2025},
  organization={IEEE}
}

@article{gnoni2017near,
  title={Near-miss management systems and observability-in-depth: Handling safety incidents and accident precursors in light of safety principles},
  author={Gnoni, Maria Grazia and Saleh, Joseph Homer},
  journal={Safety science},
  volume={91},
  pages={154--167},
  year={2017},
  publisher={Elsevier}
}

@article{phuong2026gdm,
  title={{GDM AI Control Roadmap}},
  author={Phuong, Mary and Jenner, Erik and Simon, Laurent and Ho, Lewis and Shah, Rohin and Farquhar, Sebastian and Coull, Scott},
  year={2026}
}

@misc{Williams_Sun_Sekhar_Carroll_Robinson_Kivlichan, 
url={https://openai.com/index/how-we-monitor-internal-coding-agents-misalignment/}, 
journal={How we monitor internal coding agents for misalignment}, 
publisher={OpenAI},
year = {2026},
author={Williams, Marcus and Sun, Hao and Sekhar, Swetha and Carroll, Micah and Robinson, David G. and Kivlichan, Ian}}

@article{stickland2025async,
  title={{Async Control: Stress-testing Asynchronous Control Measures for LLM Agents}},
  author={Stickland, Asa Cooper and Michelfeit, Jan and Mani, Arathi and Griffin, Charlie and Matthews, Ollie and Korbak, Tomek and Inglis, Rogan and Makins, Oliver and Cooney, Alan},
  journal={arXiv preprint arXiv:2512.13526},
  year={2025}
}

@misc{aiaaic_classifications,
  author       = {{AIAAIC}},
  title        = {{Classifications and Definitions}},
  howpublished = {AIAAIC Repository},
  year         = {2025},
  note         = {{Accessed June 28, 2026}},
  url          = {https://www.aiaaic.org/aiaaic-repository/classifications-and-definitions}
}

@misc{california_sb53,
  author = {{California State Legislature}},
  title  = {{Senate Bill No. 53}},
  year   = {2025},
  note         = {{Accessed June 28, 2026}},
  url    = {https://leginfo.legislature.ca.gov/faces/billTextClient.xhtml?bill_id=202520260SB53}
}

@misc{ny_raise_act,
  author = {{New York State Assembly}},
  title  = {{Assembly Bill A09449}},
  year   = {2026},
  note         = {{Accessed June 28, 2026}},
  url    = {https://nyassembly.gov/leg/?bn=A09449&term=2025&Text=Y}
}

@article{slattery2026ai,
  title={{The AI risk repository: A meta-review, database, and taxonomy of risks from artificial intelligence}},
  author={Slattery, Peter and Saeri, Alexander K and Grundy, Emily AC and Graham, Jess and Noetel, Michael and Uuk, Risto and Dao, James and Pour, Soroush and Casper, Stephen and Thompson, Neil},
  journal={Patterns},
  year={2026},
  publisher={Elsevier}
}

@inproceedings{yampolskiy2016taxonomy,
  title={{Taxonomy of Pathways to Dangerous Artificial Intelligence.}},
  author={Yampolskiy, Roman V},
  booktitle={AAAI Workshop: AI, Ethics, and Society},
  pages={143--148},
  year={2016}
}

@misc{mit_ai_incident_tracker,
  author       = {{MIT AI Risk Repository}},
  title        = {{AI Incident Tracker}},
  howpublished = {MIT AI Risk Initiative, MIT FutureTech},
  year         = {2026},
  note         = {Accessed June 28, 2026},
  url          = {https://airisk.mit.edu/ai-incident-tracker}
}

@misc{csetv1,
  author       = {{AIID}},
  title        = {{CSETv1 Charts}},
  year         = {2026},
  note         = {Accessed June 28, 2026},
  url          = {https://incidentdatabase.ai/taxonomies/csetv1/}
}

@misc{mcgregor2021taxonomy,
  author       = {McGregor, Sean},
  title        = {{The First Taxonomy of AI Incidents}},
  howpublished = {AIID (blog)},
  year         = {2021},
  month        = jul,
  note         = {Posted July 8, 2021. Accessed June 28, 2026},
  url          = {https://incidentdatabase.ai/blog/the-first-taxonomy-of-ai-incidents/}
}

@misc{aiid_taxonomies,
  author       = {{AIID}},
  title        = {{List of Taxonomies}},
  year         = {2026},
  note         = {Accessed June 28, 2026},
  url          = {https://incidentdatabase.ai/taxonomies/}
}

@article{shao2025your,
  title={Your agent may misevolve: Emergent risks in self-evolving llm agents},
  author={Shao, Shuai and Ren, Qihan and Qian, Chen and Wei, Boyi and Guo, Dadi and Yang, Jingyi and Song, Xinhao and Zhang, Linfeng and Zhang, Weinan and Liu, Dongrui and others},
  journal={arXiv preprint arXiv:2509.26354},
  year={2025}
}

@inproceedings{wei2026designing,
  title={Designing incident reporting systems for harms from general-purpose AI},
  author={Wei, Kevin and Heim, Lennart},
  booktitle={Proceedings of the AAAI Conference on Artificial Intelligence},
  volume={40},
  number={44},
  pages={38016--38029},
  year={2026}
}
\bibliographystyle{icml2024}

\newpage
\appendix
\onecolumn
\section{Corporate Monitoring Policies}
\label{sec:policies}

Corporate monitoring policies provide insights into how leading AI developers interpret their post-deployment monitoring responsibilities. We analysed publicly available AI safety and security frameworks to gain an understanding of how these organisations publicly articulate their post-deployment monitoring responsibilities. This analysis does not reflect the full scope of their internal practices. 

Across the surveyed materials, several common patterns emerge in how organisations frame and operationalise monitoring. User-driven reporting channels and anonymous employee escalation pathways provide avenues for people engaging with the system to report issues, abuse, or incidents. Some companies offer bug bounty programmes, which actively incentivise user reports. Many frameworks also reference logging practices, including prompt and invocation logs, infrastructure-change logs, and classifiers for misuse detection. About half of the policies reviewed indicated a dedicated incident response and investigation framework, including internal oversight teams responsible for reviewing alerts and conducting investigations. Some organisations described red-teaming and post-launch evaluations to identify regressions or new risk patterns. 

Our analysis reveals substantial variation in scope, specificity, and operational detail. 

Below, we summarise how organisations are currently interpreting and operationalising monitoring obligations.

\begin{table}[H]
\centering
\caption{Publicly disclosed post-deployment monitoring practices across
frontier AI developers. Derived from each organisation's publicly available
AI safety and security framework; does not reflect the full scope of
internal practices.}
\label{tab:monitoring-policies}
\footnotesize
\setlength{\tabcolsep}{4pt}
\renewcommand{\arraystretch}{1.2}
\begin{tabular}{@{}l cccccccc @{}}
\toprule
\textbf{Vendor}
  & \textbf{Reporting}
  & \textbf{Bug}
  & \textbf{User}
  & \textbf{Output}
  & \textbf{Escalation}
  & \textbf{Incident}
  & \textbf{Red}
  & \textbf{Downstream} \\
  & \textbf{channels}
  & \textbf{bounties}
  & \textbf{monitoring}
  & \textbf{monitoring}
  & \textbf{\& whistle-}
  & \textbf{response}
  & \textbf{teaming}
  & \textbf{attribution} \\
  &
  &
  &
  &
  & \textbf{blowing}
  &
  &
  & \\
\midrule
Amazon (AWS)    & \checkmark & \checkmark & \checkmark & \checkmark & --         & \checkmark & \checkmark & --         \\
Anthropic       & \checkmark & \checkmark & --         & \checkmark & \checkmark & \checkmark & $\bullet$  & --         \\
Cohere          & --         & \checkmark & \checkmark & $\bullet$  & --         & $\bullet$  & \checkmark & --         \\
G42             & \checkmark & --         & \checkmark & \checkmark & \checkmark & \checkmark & $\bullet$  & --         \\
Google DeepMind & --         & \checkmark & \checkmark & $\bullet$  & --         & $\bullet$  & $\bullet$  & --         \\
Magic           & --         & \checkmark & --         & $\bullet$  & $\bullet$  & $\bullet$  & $\bullet$  & --         \\
Meta            & --         & $\bullet$  & --         & $\bullet$  & --         & --         & $\bullet$  & --         \\
Microsoft       & \checkmark & \checkmark & \checkmark & \checkmark & $\bullet$  & \checkmark & $\bullet$  & \checkmark \\
NVIDIA          & \checkmark & \checkmark & --         & \checkmark & --         & \checkmark & \checkmark & \checkmark \\
OpenAI          & \checkmark & \checkmark & \checkmark & $\bullet$  & \checkmark & $\bullet$  & \checkmark & --         \\
xAI             & --         & $\bullet$  & \checkmark & \checkmark & \checkmark & $\bullet$  & \checkmark & --         \\
\bottomrule
\end{tabular}

\vspace{4pt}
{\footnotesize\raggedright
\checkmark\ explicitly present;\ \
--\ no mention;\ \
$\bullet$\ ambiguous or high-level commitment without operational detail.\par}
\end{table}

\begin{small}
\begin{longtable}{L{2.2cm} L{12cm}}
\toprule
\textbf{Vendor} & \textbf{Monitoring Policies} \\
\midrule
\endfirsthead
 
\toprule
\textbf{Vendor} & \textbf{Monitoring Policies} \\
\midrule
\endhead
 
\midrule
\multicolumn{2}{r}{\textit{Continued on next page}} \\
\endfoot
 
\bottomrule
\endlastfoot
 
Amazon (AWS) &
Amazon outlines mechanisms for external vulnerability reporting, internal threat detection, and structured incident response. A dedicated Cyber Threat Intelligence team continually monitors and tracks down advanced threat actor groups. Amazon maintains continued engagement with and investments in external security research, including bug bounty programs, academic research investments, red teaming networks, and coordinated vulnerability disclosure programs that encourage and reward security experts to partner with them for research and development.
 
Amazon employs output moderation systems to ensure that generated content adheres to its Amazon Responsible AI objectives by blocking or safely modifying violating inputs and outputs. Their Security Operations Centres provide 24/7 centralised global support. There are also dedicated Incident Response Protocols for incident escalation and response pathways in the event of reported AI safety incidents. \\
\midrule
 
Anthropic &
Anthropic Responsible Scaling Policy lists monitoring processes including responses to jailbreak bounties, conducting historical analysis or background monitoring, and any necessary retention of logs for these activities. Model capabilities are consistently evaluated and a dedicated Responsible Scaling Officer is accountable for the implementation of any interim processes that may be required if capability evaluations exceed safety thresholds. Anthropic has a Responsible Disclosure policy and a safety issue reporting platform where users may email issues to a dedicated user safety email address. They also operate private bug bounty programs through HackerOne, including programs for identifying vulnerabilities in their models and security vulnerabilities.
 
Anthropic employees can report AI safety-related concerns through three main channels, including an anonymous channel for reporting potential violations of their AI safety commitments. The Responsible Scaling Policy states that red-teaming is conducted pre-deployment. \\
\midrule
 
Cohere &
Cohere has a Secure AI Frontier Model Framework, where they describe continuous monitoring processes including unexpected post-deployment usage patterns and manual and automatic security risk identification processes. They also have ``Secure Product Lifecycle'' controls that include security risk assessments, penetration testing, and bug bounty programs. They incentivise third-party vulnerability discovery via clear protections for legitimate research practices in their Responsible Disclosure Policy. Cohere has an incident response plan whereby incidents are identified, tracked, and resolved, although operational details for this plan are not provided. Cohere conducts evaluations and testing for general performance, safety, and security throughout the lifecycle and includes red-teaming, evaluations with academic and industry benchmarks, and internal evaluations. \\
\midrule
 
G42 &
G42 has a Frontier AI safety Framework, which describes model robustness testing and asynchronous monitoring protocols. Responsibilities of their dedicated Frontier AI Governance Board include developing a comprehensive incident response plan that outlines the steps to be taken in the event of non-compliance. They outline an incident detection strategy that leverages automated mechanisms and human review. Incident reporting channels include designated pathways for users to report instances of concerning or harmful behavior in violation of company policy to relevant G42 Personnel. \\
\midrule
 
Google DeepMind &
Google's Frontier Safety Framework details post-market monitoring strategies based on model capabilities, which are evaluated proactively and throughout the entire lifecycle of the model. Deployment mitigations include safety fine-tuning and monitoring and response. If capability evaluations reach a threshold where model reasoning ability is determined to have the potential to undermine human control, additional monitoring processes such as automated chain-of-thought monitoring of the model's reasoning may be implemented. Gemini's API policy documents specify the use of automated and manual mechanisms to detect violations of the API usage policy. An AI Vulnerability Rewards Program functions as a responsible disclosure platform, and Google commits to developing a suite of safeguards which may include safety post-training, monitoring and analysis, account moderation, user verification, and bug bounties. \\
\midrule
 
Magic &
Magic has an AGI Readiness Policy where they state that they continuously monitor their AI systems to evaluate if their models have reached dangerous capability frontiers. Output safety classifiers are used for output monitoring, and automated detection is applied for internal usage within Magic. In cases where risks for threat models pass a set threshold, safety measures including delaying or pausing development are specified. These threat mitigations include both security measures such as training and compartmentalisation, as well as deployment mitigations such as harm refusal. Magic has a Vulnerability Disclosure Policy that allows security researchers to submit suspected vulnerabilities via a web form. \\
\midrule
 
Meta &
Meta's Frontier AI Framework includes pre-deployment AI risk assessments and threat modelling exercises that identify potential risks associated with frontier AI, enable the development of risk mitigation strategies, and inform model deployment plans. \\
\midrule
 
Microsoft &
Microsoft implements post-deployment tools and processes for ongoing monitoring, user feedback channels, incident response, and iterative improvements to its risk mitigation stack. Product teams across Microsoft are required to put in place repeatable processes to collect user feedback and to triage and address issues that arise after the release of an AI system. Microsoft also has a Coordinated Vulnerability Disclosure program, where researchers may disclose issues with AI systems. The program includes a Bug Bounty Program to incentivise users to report significant security issues. At the platform level, safety measures include content classifiers that block potentially harmful user inputs and AI-generated content. Microsoft uses a unified API to detect and block jailbreak patterns in user inputs and indirect prompt injection attacks.
 
A Microsoft Security Response Center researcher portal also accepts reports for potential incidents. These reports may be anonymous. Microsoft details an incident response plan that includes expanding the capacity of specialised roles like crisis managers, forensic investigators, and communications managers. An AI Red Teaming Agent is made available to customers, who may use this to simulate attack techniques and generate red teaming reports that help track risk mitigation improvements throughout the AI development life cycle.
 
To support downstream monitoring of AI tools, services, and components, Microsoft released `Azure AI Foundry Observability' in May 2025, which offers continuous monitoring and evaluation of both AI applications and agentic AI systems in production. \\
\midrule
 
NVIDIA &
NVIDIA's Frontier AI risk Assessment framework outlines strategies for lowering hazard duration, decreasing hazard onset speed. It states that NVIDIA models have watermarks embedded at generation-time that enable downstream detection and attribution of AI-produced outputs, which provides a verifiable origin signal for both end-users and automated scanning tools. NVIDIA also has a Vulnerability Disclosure program where security vulnerabilities can be submitted. To address evolving threats and vulnerabilities, red teaming activities are used in conjunction with public benchmarks. \\
\midrule
 
OpenAI &
OpenAI's preparedness framework lists safeguards against harm, including usage monitoring to stop or catch adversarial users, adversarial testing, and red-teaming. A monitor and incident response plan is cited, although operational details are not provided. OpenAI also has a coordinated vulnerability disclosure policy including a bug bounty program and an incident reporting platform. \\
\midrule
 
xAI &
xAI uses real-time monitoring, telemetry and alerting of threshold breaches via internal tooling. Models are equipped with the ability to scrutinise user behaviour and identify bad actors. Classifiers are applied to user inputs to verify safety when a model is queried regarding weapons of mass destruction or cyberterrorism. xAI also has a responsible disclosure program that includes bug bounties to encourage external parties to report security issues. Employees may anonymously report concerns about non-adherence to their risk management framework. \\
 
\end{longtable}
\end{small}

\section{Monitoring principles}
\label{sec:monitoringprinciples}

Post-deployment incident monitoring techniques should account for both the informational requirements of the reporting, forecasting, and learning stages that follow, as well as the challenges outlined above. Accordingly, the principles detailed below are designed to optimise these activities: 

\begin{itemize}
    \item \textbf{Continuous:} Monitoring should be maintained throughout the system’s operational life cycle and proportionate to its risk profile \cite{szadeczky2025risk} with the aim of minimising both the Mean Time to Detection (MTTD) \cite{johnson2024developing} and the likelihood of undetected incidents. 
    \item \textbf{Calibrated:} Alert mechanisms should be tuned to maintain a balanced trade-off between false positives and false negatives. Calibration should be periodically reviewed to prevent alert fatigue and ensure that monitoring processes remain accurate and timely, with MTTD targets proportionate to system and context risk \cite{o2023deployment, tariq2025alert}.
    \item \textbf{Traceable:} Monitoring data should be comprehensive enough to trace the sequence of system states, inputs, outputs, and interactions that led to an incident, enabling independent verification of what occurred and supporting later causal analysis \cite{euaiact}. Traceability should connect technical evidence such as system event logs \cite{yampolskiy2025monitorability, paeth2024lessons} and logs of user or agent interactions \cite{ezell2025incident} with organisational records documenting existing safeguards and policies \cite{dixon2025ai} and updates to risk and safety case evaluations \cite{buhl2024safety}.
    \item \textbf{Impact inclusive:} Once an incident is detected, monitoring should ensure the systematic capture and preservation of observable impact-relevant evidence, including signals from which scope, frequency, duration, and severity can later be derived \cite{vei2025ai}. This should include the number of users or systems affected, how long the harm or disruption persisted, and how severe the consequences were across technical, organisational, or social dimensions \cite{mylius2023aiharmseverity, hoffmann2023addingstructureaiharm}. Impact data may be drawn from telemetry, user feedback, impact assessments, or incident reports. 
    \item \textbf{Privacy-preserving defaults:} Monitoring that involves user data requires the informed consent of users \cite{stein2024role} or the implementation of rigorous anonymisation techniques such as pseudonymisation or aggregation methods \cite{bluemke2023exploring}. Minimal sensitive data should be processed and cyber threat evaluations should be conducted to identify risks of re-identification or leakage \cite{shahriar2023survey}. Information flows, including how sensitive data moves within and between systems, for what purposes, and for how long, should be documented and auditable, ensuring that privacy is maintained through controlled and transparent data movement \cite{bluemke2023exploring}.
\end{itemize}

These principles and their associated guidelines (Appendix \ref{sec:guidelines}) provide a structured foundation for post-deployment incident monitoring, however, they interact in ways that create practical trade-offs. Privacy-preserving data minimisation can limit the granularity of logs required for full traceability, while achieving comprehensive traceability or implementing continuous, proactive signal collection may require collecting data that increases privacy risk or retention burdens. Calibration choices may also affect impact-relevant signals, particularly when it is uncertain which metrics will later prove important for assessing the severity of harm. As a result, optimising each principle independently may compromise overall monitoring effectiveness. Organisations must balance trade-offs between these principles in light of system risk, technical feasibility and legal obligations, recognising that monitoring is a complex system design problem rather than a checklist of isolated requirements.

\section{Reporting principles}
\label{sec:reportingprinciples}

We propose five principles for AI incident reports:

\begin{itemize}
    \item \textbf{Iterative:} Reporters should identify whether a report is an initial, follow-up or final report. Receiving bodies should assign an ID to initial reports, which reporters can reference in follow-up and final reports to link them. Reports should be automatically timestamped at submission. This reflects the reality that information is uncovered gradually while supporting traceability. 
    \item \textbf{Pragmatic:} Initial reports should request information that reporters can reasonably provide: system identification, usage context, core harm types, approximate harm quantification, and suspected causes. Detailed technical information and comprehensive causal analysis can be provided in follow-up reports if necessary.
    \item \textbf{Epistemically transparent:} Initial reports should accommodate uncertainty, requiring reporters to specify whether their inputs are known, estimated or unknown. 
    \item \textbf{Unambiguous:} Questions should be worded precisely and structured to help reporters identify relevant information. For example, rather than asking about “likely causality,” questions can suggest factors to consider, such as data, model, system, usage context, or governance failures.
    \item \textbf{Analysable:} Reports should be exportable in machine-readable formats so their contents can be annotated using more granular taxonomies, enabling comparison and learning across incidents.
\end{itemize}

\clearpage
\section{Monitoring Guidelines} 
\label{sec:guidelines}

\begin{small}
\begin{longtable}{L{1.8cm} L{3.8cm} L{5.8cm} L{2.2cm}}
\toprule
\textbf{Principle} & \textbf{Guideline} & \textbf{Operationalisation Details} & \textbf{Guidelines Derived From} \\
\midrule
\endfirsthead
 
\toprule
\textbf{Principle} & \textbf{Guideline} & \textbf{Operationalisation Details} & \textbf{Guidelines Derived From} \\
\midrule
\endhead
 
\midrule
\multicolumn{4}{r}{\textit{Continued on next page}} \\
\endfoot
 
\bottomrule
\endlastfoot
 
M1. Continuous
& M1.1 Monitoring should be ongoing and maintained throughout all post-deployment stages of the AI system life cycle.
& Monitoring should occur while the AI system is in operation, from deployment to decommissioning. Choice of monitoring modalities (e.g., real-time, periodic, or event-triggered) should be risk-based and system architecture-dependent (see M1.2).
& \citet{iso_iec_22989_2022} \S6.2.6--6.2.7; \citet{iso_iec_23053_2022} \S8.7 \\
\cmidrule(l){2-4}
 
M1. Continuous
& M1.2 Continuous monitoring modalities should reflect and be proportional to the risk profile of the system and context of operation.
& Risk-based monitoring regimes should be implemented, such that AI systems with higher risk classifications are subject to more frequent, more detailed, and more comprehensive monitoring. Monitoring requirements associated with a system's assigned risk tier and risk treatment plan should be explicitly documented, periodically reviewed, and updated following model updates, changes in deployment context, or the emergence of new risk signals.
& \citet{nist2023rmf} \S6.4; \citet{euaiact} Art.~72; \citet{iso_iec_23894_2023} \S6.7 \\
\cmidrule(l){2-4}
 
M1. Continuous
& M1.3 Continuous monitoring mechanisms should be designed with the aim to minimise Mean Time to Detection (MTTD).
& Organisations should define MTTD targets proportional to the objectives and risks of the system and review them periodically. Automated monitoring mechanisms should flag distribution shifts, usage patterns that indicate anomalous outputs, or deviations from expected behaviour to minimise the likelihood of undetected incident onset. Monitoring infrastructure should support low-latency detection through automated alerts and structured escalation pathways. Incident categories without a clear onset (e.g., intangible, distributed, or long-horizon harms) may be more challenging to detect, and therefore may have higher MTTDs, but organisations should still aim to minimise the time to detection for these incidents. Escalation rules (such as external audits) for when MTTD targets are exceeded should be set.
& \citet{johnson2024developing, o2023deployment, hoffmann2023addingstructureaiharm} \\
\midrule
 
M2. Calibrated
& M2.1 Tolerance bands for false positives and false negatives should be established and documented.
& Define thresholds for false-positive and false-negative tolerance. Record the number of false-positive and false-negative alerts.
& \citet{vinnakotacreating, tariq2025alert} \\
\cmidrule(l){2-4}
 
M2. Calibrated
& M2.2 Monitoring thresholds should be recalibrated on a regular schedule and after major system changes.
& Detection thresholds, statistical baselines, and alert triggers should be recalibrated at defined intervals (e.g., monthly/quarterly) and after events such as model updates, safety-filter changes, or significant distribution shifts. Recalibration decisions should be logged and validated.
& \\
\cmidrule(l){2-4}
 
M2. Calibrated
& M2.3 Multi-layered detection combining automated and human signals should be implemented.
& Monitoring should integrate automated anomaly detection, usage-pattern monitors, and structured human review for high-severity alerts. User reports, trusted researcher inputs, and partner feedback should serve as additional layers for capturing signals that automated systems may miss.
& \\
\cmidrule(l){2-4}
 
M2. Calibrated
& M2.4 Alert volume should be managed to prevent alert fatigue without compromising safety.
& Monitoring systems should implement alert prioritisation, deduplication, and routing rules to reduce noise and prevent operational overload. Suppression logic must be transparent, auditable, and prohibited from hiding alerts in high-severity categories or suppressing new anomaly types.
& \citet{o2023deployment, tariq2025alert} \\
\midrule
 
M3. Traceable
& M3.1 Runtime behavioural data necessary to understand the sequence of events leading to an incident should be recorded.
& Monitoring should capture inputs, outputs, model version identifiers, configuration parameters, and relevant runtime metadata with the goal of tracing system behaviour that triggered or otherwise contributed to the incident.
& \citet{anderljung2023frontier}; \citet{ezell2025incident}; \citet{ISO_IEC_42001_2023} \S B.6.2.6; \citet{eu_gpai_code} Ch.~3 \\
\cmidrule(l){2-4}
 
M3. Traceable
& M3.2 A complete and time-stamped record of lifecycle and configuration changes should be maintained.
& Logs should include model updates, fine-tuning, safety-layer modifications, tool-access changes, and deployment approvals.
& \\
\cmidrule(l){2-4}
 
M3. Traceable
& M3.3 Relevant external information sources used to identify or corroborate incidents should be monitored and preserved.
& Organisations should track user reports, news coverage, public disclosures, or research publications that provide details relating to the occurrence and possible causes of an incident.
& \citet{iso_iec_27035-1_2023} ; \citet{eu_gpai_code} Ch.~3 Commitment~9 \\
\cmidrule(l){2-4}
 
M3. Traceable
& M3.4 Post-deployment technical and organisational changes relevant to the incident should be captured.
& Technical and organisational changes that affect system operation or oversight, including updates to safeguards, operating procedures, oversight structures, and risk controls, should be documented when they occur.
& \citet{dixon2025ai} \\
\cmidrule(l){2-4}
 
M3. Traceable
& M3.5 Monitoring records should be tamper-evident, access-controlled, and auditable.
& Logs should be stored in formats that make unauthorised modification detectable. Access to logs should follow role-based controls and generate audit trails.
& \\
\midrule
 
M4. Impact-inclusive
& M4.1 Observable indicators of an incident's scope, frequency, duration, and severity should be recorded.
& Once an incident is detected, monitoring should capture measurable indicators of how widely the harm extends (scope), how often it occurs (frequency), how long it persists (duration), and the magnitude of its effects (severity). Indicators may include the number of affected users or systems, durations of service disruption, error rates, or the extent of policy or safeguard failures.
& \citet{scheuerman2021framework, hoffmann2023addingstructureaiharm, mylius2023aiharmseverity} \\
\cmidrule(l){2-4}
 
M4. Impact-inclusive
& M4.2 Impact evidence should be collected through ecosystem monitoring and user-facing channels.
& Organisations should monitor the ecosystem in which the AI system is deployed for issues and maintain awareness of new AI research findings and techniques. Organisations should provide channels for collecting user feedback or reports related to the impact of harm arising from an incident.
& \citet{iso_iec_23894_2023}; \citet{eu_gpai_code} Ch.~3 \\
\cmidrule(l){2-4}
 
M4. Impact-inclusive
& M4.3 Impact-relevant signals should be timestamped to track incident progression.
& Monitoring systems should preserve when impact indicators emerged, escalated, and resolved, enabling reconstruction of how the harm developed over time. For diffused or long-horizon harms where precise timing is difficult to establish, relevant indicators should still be recorded where feasible, and the absence of temporal precision should be documented.
& \\
\midrule
 
M5. Privacy-preserving defaults
& M5.1 Lawful and privacy-protective bases for collecting monitoring-relevant data should be used.
& Monitoring should only collect personal data under a lawful basis such as meaningful consent, contractual necessity, legitimate interests, or compliance with regulatory obligations, depending on the deployment context. The purpose of collecting monitoring data must be clearly specified and limited to incident detection and documentation. Data-minimisation checks should be performed before designing a data collection plan.
& \citet{stein2024role} \\
\cmidrule(l){2-4}
 
M5. Privacy-preserving defaults
& M5.2 The collection and retention of sensitive or identifiable data should be minimised.
& Monitoring should only collect data strictly necessary for detecting and documenting incidents. Sensitive categories of data (e.g., biometrics, health data) should be avoided unless essential.
& \\
\cmidrule(l){2-4}
 
M5. Privacy-preserving defaults
& M5.3 Rigorous anonymisation or pseudonymisation should be employed.
& Where feasible, logs should use privacy-preserving techniques such as pseudonymisation, or aggregation.
& \citet{feretzakis2024trustworthy} \\
\cmidrule(l){2-4}
 
M5. Privacy-preserving defaults
& M5.4 Risks of re-identification and data leakage should be evaluated and mitigated.
& Organisations should conduct cyber-threat evaluations or privacy risk assessments to identify potential re-identification pathways or leakage vectors in monitoring data, including through linkage attacks, model inversion risks, or correlation across logs. Privacy impact assessments or risk reviews should be performed periodically.
& \citet{shahriar2023survey, xin2025false, iso_iec_23894_2023} \\
\cmidrule(l){2-4}
 
M5. Privacy-preserving defaults
& M5.5 Information flows related to monitoring data should be documented and audited.
& Data-flow diagrams detailing how sensitive data moves within and between systems, for what purposes, and for how long, should be documented and auditable. Retention and deletion policies must balance forensic integrity with data protection requirements.
& \citet{bluemke2023exploring}\\
 
\end{longtable}
\end{small}

\section{Reporting Template}
\label{sec:template}

The following template operationalises the principles discussed in Section \ref{sec:reportingprinciples}.

The template assumes reports are submitted to an external body that assigns an ID to a report series, enabling initial, follow-up, and final reports to be linked over time (Iterative). It accommodates incomplete or uncertain information, allowing reporters to specify whether inputs are known, estimated, or unknown (Epistemically transparent), and add detail in subsequent reports (Pragmatic). Explanatory text in italics clarifies what information is being requested (Unambiguous).

Fields relating to implicated systems are intended to be repeatable where reporters are aware that multiple systems contributed to an incident. In such cases, differential requirements may apply to main contributing systems and secondary systems.

\medskip
 
\newcommand{\formfield}{\rule{\linewidth}{0.4pt}}
\newcommand{\reqmark}{\textbf{*}}
\newcommand{\formsection}[1]{\cellcolor{gray!15}\textbf{#1} \\[2pt]}
\newcommand{\formhelp}[1]{\quad\textit{\small #1} \\[2pt]}
\newcommand{\formoptions}[1]{\quad\small #1 \\[2pt]}
\newcommand{\formentry}[1]{\quad #1 \\[1pt]}
\newcommand{\formfieldrow}{\quad\rule{0.9\textwidth}{0.4pt} \\[4pt]}
\newcommand{\kepistemic}{$\square$~Known \quad $\square$~Estimated \quad $\square$~Unknown}
 
\renewcommand{\arraystretch}{1.3}
 
\begin{small}
\setlength{\LTleft}{0pt}
\setlength{\LTright}{0pt}
\begin{longtable}{@{} p{0.96\textwidth} @{}}
 
\toprule
\cellcolor{gray!25}\large\textbf{Initial AND/OR Follow-up Report} \\
\midrule
\endfirsthead
 
\toprule
\cellcolor{gray!25}\large\textbf{AI Incident Report (continued)} \\
\midrule
\endhead
 
\midrule
\hfill\textit{Continued on next page} \\
\endfoot
 
\bottomrule
\endlastfoot
 
\formsection{Report Metadata}
\formentry{\reqmark~Report type} 
\formoptions{$\square$~Initial \quad $\square$~Follow-up \quad $\square$~Final}
\formentry{Incident ID}
\formhelp{Leave blank if Initial Report.}
\formfieldrow
\formentry{\reqmark~Date of Submission}
\formhelp{Auto-populated.}
\formentry{YYYY-MM-DD HH:MM}
\midrule
 
\formsection{Reporter Information}
\formentry{\reqmark~Surname}
\formfieldrow
\formentry{\reqmark~First name}
\formfieldrow
\formentry{\reqmark~Affiliation}
\formhelp{Enter affiliation or select ``No Affiliation.''}
\formfieldrow
\formentry{\reqmark~Role}
\formhelp{Select all that apply.}
\formoptions{$\square$~Developer \quad $\square$~Deployer \quad $\square$~User \quad $\square$~Third Party \quad $\square$~Other (specify)}
\formentry{Contact details for follow-up}
\formfieldrow
\midrule
 
\formsection{Timeline}
\formentry{\reqmark~Start date}
\formhelp{If known or estimated, enter the date when the incident first occurred (as precisely as possible).}
\formfieldrow
\formoptions{\kepistemic}
\formentry{\reqmark~Detection date}
\formhelp{If known or estimated, enter the date when the incident was first detected (as precisely as possible).}
\formfieldrow
\formoptions{\kepistemic}
\formentry{\reqmark~End date}
\formhelp{If known or estimated, enter the date when the system(s) believed to have contributed to the incident was/were restored to normal functioning (as precisely as possible).}
\formfieldrow
\formoptions{\kepistemic}
\midrule
 
\formsection{Incident Description}
\formentry{Describe what the system did or failed to do.}
\formhelp{Describe observable system behaviour. Avoid causal explanations.}
\formfieldrow
\midrule
 
\formsection{Implicated System(s)}
\formhelp{Provide the following information for each system known to the reporter whose behaviour may have contributed to the incident.}
\formentry{System's role in the incident}
\formoptions{$\square$~Main contributing system \quad $\square$~Other contributing system (specify)}
\formentry{\reqmark~Model name}
\formfieldrow
\formentry{\reqmark~Model version}
\formfieldrow
\formentry{Release date}
\formfieldrow
\formentry{\reqmark~Intended use}
\formhelp{Select all that apply.}
\formoptions{$\square$~Facial recognition \quad $\square$~Content moderation \quad $\square$~Medical diagnosis}
\formoptions{$\square$~Hiring/recruitment \quad $\square$~Financial services \quad $\square$~Autonomous vehicles}
\formoptions{$\square$~Conversational AI \quad $\square$~Content generation \quad $\square$~Other (specify)}
\formentry{Actual use}
\formhelp{Describe how the system was being used at the time of the incident.}
\formfieldrow
\formentry{Deployment configuration}
\formhelp{Describe relevant aspects of the system's deployment or operational setting. Include, where applicable: (1) the system's level of autonomy; (2) integration with external tools or APIs.}
\formfieldrow
\formentry{Known limitations or safety measures}
\formhelp{Describe any known limitations, safeguards, or mitigations relevant to the incident.}
\formfieldrow
\formentry{Model type and architecture}
\formhelp{Describe the model type (e.g.\ large language model, image classifier) and architecture (e.g.\ transformer, CNN, diffusion model).}
\formfieldrow
\formentry{Interaction with other systems}
\formhelp{Describe known interactions with other AI or non-AI systems relevant to the incident.}
\formfieldrow
\midrule
 
\formsection{Impact}
\formentry{\reqmark~Impacted parties}
\formhelp{Describe who (e.g.\ users, non-users) or what was affected by the incident.}
\formfieldrow
\formentry{\reqmark~Nature of impact}
\formhelp{Describe how they were affected.}
\formfieldrow
\formentry{\reqmark~Harm types}
\formhelp{Select all that apply.}
\formoptions{$\square$~Physical harm \quad $\square$~Psychological harm \quad $\square$~Financial harm}
\formoptions{$\square$~Environmental harm \quad $\square$~Discrimination/bias \quad $\square$~Privacy violation}
\formoptions{$\square$~Reputational harm \quad $\square$~Other (specify)}
\formentry{Number of users impacted}
\formfieldrow
\formoptions{\kepistemic \quad Additional detail: \rule{3cm}{0.4pt}}
\formentry{Number of non-users impacted}
\formfieldrow
\formoptions{\kepistemic \quad Additional detail: \rule{3cm}{0.4pt}}
\formentry{Economic damage}
\formfieldrow
\formoptions{\kepistemic \quad Additional detail: \rule{3cm}{0.4pt}}
\midrule
 
\formsection{Causality}
\formentry{\reqmark~Describe suspected contributing factors to the incident.}
\formhelp{Include, where applicable: (1) technical factors related to the system, data, model, or deployment; (2) non-technical factors related to human use, organisational, or governance context.}
\formfieldrow
\formentry{\reqmark~Suspected contributing factors}
\formhelp{Select all that apply.}
\formoptions{$\square$~Data issue \quad $\square$~Model issue \quad $\square$~System issue \quad $\square$~Deployment context}
\formoptions{$\square$~User behaviour \quad $\square$~Organisational context \quad $\square$~Governance or policy gaps}
\formoptions{$\square$~External factors \quad $\square$~Unknown}
 
\end{longtable}
\end{small}

\end{document}